\documentclass[10pt,aps,prc,onecolumn,notitlepage,superscriptaddress,floatfix,nofootinbib,showkeys]{revtex4-1}
\usepackage{amsmath,amssymb,mathrsfs}
\usepackage{bm}
\usepackage{ascmac}
\usepackage{geometry}
\usepackage{graphicx}

\begin{document}

\title{
TDHF Theory and Its Extensions for the Multinucleon Transfer Reaction:\\
a Mini Review
}

\author{Kazuyuki Sekizawa}
\email[]{sekizawa@phys.sc.niigata-u.ac.jp}
\affiliation{Center for Transdisciplinary Research, Institute for Research Promotion, Niigata University, Niigata 950-2181, Japan}

\date{February 5, 2019}

\begin{abstract}
Time-dependent Hartree-Fock (TDHF) theory has been a powerful tool in describing a variety of complex
nuclear dynamics microscopically without empirical parameters. In this contribution, recent advances in nuclear
dynamics studies with TDHF and its extensions are briefly reviewed, along the line with the study of multinucleon
transfer (MNT) reactions. The latter lies at the core of this Research Topic, whose application for production of
extremely neutron-rich nuclei has been extensively discussed in recent years. Having in mind the ongoing
theoretical developments, it is envisaged how microscopic theories may contribute to the future MNT study.
\end{abstract}


\maketitle

\section{Introduction}

The MNT reaction may be a promising means to produce yet-unknown neutron-rich unstable nuclei,
whose production is difficult by other methods. This possibility makes the MNT study fascinating and important.
For instance, besides the fundamental interest in nuclear structure such as shell evolution \cite{Otsuka(2018)}
and shape transitions \cite{Cejnar(2010),Heyde(2011)}, properties of extremely neutron-rich nuclei are crucial
to fully understand the detailed scenario of the $r$-process nucleosynthesis \cite{r-process}. Needless to say, the historic detection
of gravitational waves from a merger of two neutron stars, GW170817 \cite{GW170817}, together with radiations
from a short $\gamma$-ray burst, GRB\,170817A \cite{GRB170817A}, followed by a kilonova \cite{kilonova}
makes it a timely and imperative task to unveil inherent properties of the nuclei far away from the stability.
Furthermore, superheavy nuclei produced so far are neutron-deficient ones, and the predicted
island of stability has not been reached yet \cite{Hofmann(2000),Oganessian(2017)}. The production of
neutron-rich superheavy nuclei in the island is highly desired, as it would provide a new stringent constraint for microscopic
theories. Therefore, the study of MNT reactions is listed as one of the key subjects at the current and future
RI beam facilities, such as RIBF (RIKEN, Japan) \cite{Sakurai(2018)}, HIRFL-CSR and HIAF (IMP, China)
\cite{HIAF}, RAON (RISP, Korea) \cite{RAON}, DRIB (FLNR, Russia), SPIRAL2 (GANIL, France) \cite{Gales(2007)},
FAIR (GSI, Germany) \cite{FAIR}, SPES (INFN, Italy) \cite{SPES}, FRIB (MSU, USA) \cite{FRIB}, and so on.

This review briefly digests recent advances of microscopic reaction theories---to stimulate new
ideas for the future study of MNT reactions. In Sec.~\ref{Sec:method}, the theoretical framework of TDHF
is succinctly recalled. In Sec.~\ref{Sec:MNT}, the current status of the MNT study with TDHF is given. In
Sec.~\ref{Sec:extensions}, recent advances of theoretical approaches are outlined, along with a discussion
on possible future applications. The article is concluded in Sec.~\ref{Sec:summary}. The readers are intended
to refer to other articles in this Research Topic for discussions on experimental as well as other theoretical studies.

\section{The TDHF theory}\label{Sec:method}

The TDHF theory was first proposed by Dirac in 1930 \cite{Dirac(1930)} and its application to nuclear
physics started in the 1970s \cite{BKN(1976),Cusson(1976),Koonin(1977),Davies(1978),Flocard(1978),
Krieger(1978),Davies(1979),Negele(review)}. The TDHF theory can be formally derived based on the
time-dependent variational principle (various derivations can be found in a recent review \cite{TDHF-review(2018)}).
In TDHF the wave function is a Slater determinant for all times, and the Pauli principle is thus
automatically ensured. The TDHF equations read:
\begin{equation}
i\hbar\frac{\partial\phi_i(\boldsymbol{r}\sigma q,t)}{\partial t} = \hat{h}[\rho(t)]\phi_i(\boldsymbol{r}\sigma q,t),
\label{Eq:TDHF}
\end{equation}
where $\phi_i(\boldsymbol{r}\sigma q,t)$ is the single-particle wave function of $i$th
nucleon at position $\boldsymbol{r}$ with spin $\sigma$ and isospin $q$. $\hat{h}[\rho(t)]=
\frac{\delta E}{\delta\rho}$ is the single-particle Hamiltonian with $\rho$ being the
one-body density. $E[\rho]$ is the total energy, which may be regarded as an energy
density functional (EDF) in the context of time-dependent density functional theory (TDDFT)\footnote{
Here a local EDF (like Skyrme) has been assumed, which makes the TDHF equations (\ref{Eq:TDHF})
local in space, as it is currently used in most of practical applications. In general, $E=\int\mathcal{E}
(\boldsymbol{r})d\boldsymbol{r}$ is composed of various local densities, $\vec\xi\equiv\{\rho,\tau,\dots\}$,
and the TDHF equations (\ref{Eq:TDHF}) should read: $i\hbar\frac{\partial\phi_i(\boldsymbol{r}\sigma q,t)}
{\partial t}=\int \sum_k \frac{\delta \mathcal{E}[\vec{\xi}(\boldsymbol{r}',t)]}{\delta \xi_k(\boldsymbol{r}',t)}
\frac{\delta\xi_k(\boldsymbol{r}',t)}{\delta\phi_i^*(\boldsymbol{r}\sigma q,t)} d\boldsymbol{r}'=
\sum_{\sigma'}\hat{h}_{\sigma\sigma'}^{(q)}(\boldsymbol{r},t)\phi_i(\boldsymbol{r}\sigma'q,t)$.
In such a case, the single-particle Hamiltonian can have spin dependence as well as differential
operators (for the explicit form, see, e.g., Refs.~\cite{Engel(1975),Dobaczewski(1995),Stevenson(2019)}).
} \cite{DFT1,DFT2,DFT3,DFT4,Nakatsukasa(review)}. The EDF is constructed so as to reproduce
static properties of finite nuclei in a wide mass region and also the basic nuclear matter properties
(see, e.g., Refs.~\cite{Bender(review),UNEDF01,UNEDF2,SeaLL1} and references therein). There
is no adjustable parameter in TDHF, once an EDF is given. The non-linearity arises because $\hat{h}[\rho(t)]$
contains the mean-field potential which depends on densities generated by all the nucleons, e.g.,
$\rho(\boldsymbol{r},t)=\sum_{i,\sigma,q}|\phi_i(\boldsymbol{r}\sigma q,t)|^2$. Such non-linear
couplings between single-particle and collective degrees of freedom give rise to the so-called one-body
dissipation mechanism (known as wall-and-window formulas \cite{Blocki(1978)}). Note that as is evident
from a derivation of TDHF from the lowest order truncation of the Bogoliubov-Green-Kirkwood-Yvon (BBGKY)
hierarchy \cite{TDHF-review(2018)}, where two-body correlations are neglected, two-body dissipations
associated with nucleon-nucleon collisions are absent in TDHF. They are, however, expected to play
a minor role at low energy due to the Pauli exclusion principle.

TDHF is notably versatile---by changing initial condition, external potential, boundary condition, etc.,
one can study a wide variety of phenomena: not only collective excitations in the linear response regime
\cite{NakatsukasaYabana(2005),UO(liner-responce),Maruhn(GDR),Ebata(2010),Fracasso(2012),Pardi(2013),
Scamps(2013)LR,Scamps(2014)LR,Ebata(2014)}, where TDHF is formally equivalent to the random phase
approximation (RPA), but also induced fission \cite{Simenel(2014),Scamps(2015),Tanimura(2015),Goddard(2015),
Goddard(2016),Scamps(2018)2}, dynamics of exotic configurations like $\alpha$-chain structure \cite{Umar(2010)},
toroidal nucleus \cite{Ichikawa(2014)1,Ichikawa(2014)2}, nuclear pasta formation \cite{Schuetrumpf(2013),
Schuetrumpf(2014),Schuetrumpf(2015)1,Schuetrumpf(2015)2}, as well as heavy-ion reactions, such as nucleon
transfer \cite{Umar(2008),Iwata(2010),Evers(2011),Umar(2017),Projection,KS_KY_MNT,KS_KY_PNP,Bidyut(2015),
KS_KY_Maruhn,KS_KY_Ni-U,KS_SH_Kazimierz,KS_GEMINI,KS_U-Sn,Bidyut(2017)}, quasifission (QF) \cite{Golabek(2009),
Kedziora(2010),Simenel(2012),Wakhle(2014),Oberacker(2014),Umar(2015)1,Hammerton(2015),Washiyama(2015),
Umar(2015)3,Umar(2016)1,Yu(2017),Guo(2018)2,Zheng(2018)}, and fusion \cite{Umar(2006),Umar(2009),Umar(2010)2,
Oberacker(2010),Umar(2012)1,Oberacker(2012),Keser(2012),Oberacker(2013),Simenel(2013)1,Umar(2014),
Steinbach(2014),Reinhard(2016),Godbey(2017),Guo(2018)3,Simenel(2007),Bourgin(2016),Vo-Phuoc(2016)},
and so on (see, Refs.~\cite{Negele(review),TDHF-review(2018),Stevenson(2019),Nakatsukasa(review),
Simenel(review),Nakatsukasa(PTEP)}, for review papers).

In the case of heavy-ion reactions, the initial wave function is composed of projectile and target
nuclei in their ground state, which is obtained by self-consistently solving the static Hartree-Fock
equations [$i\hbar\partial/\partial t$\,$\rightarrow$\,$\varepsilon_i$ in Eq.~(\ref{Eq:TDHF})].
Those wave functions are placed in a computational box, with a sufficiently large relative distance,
boosted with a proper relative momentum. The time evolution according to Eq.~(\ref{Eq:TDHF})
then allows us to follow reaction dynamics in real space and real time. As the theory deals with the
single-particle wave functions of nucleons, not only dynamic effects such as nucleon transfer and
internal excitations, but also structural effects such as static/dynamic shell effects and shape evolution
are naturally incorporated into the description. The spin-orbit coupling is known to play an important
role in energy dissipation processes in heavy-ion reactions \cite{Umar(1986),Reinhard(1988),Umar(1989),
Dai(2014)2}. The effects of the tensor term on nuclear dynamics were also investigated recently
\cite{Guo(2018)3,Iwata(2011),Dai(2014)1,Stevenson(2016),Guo(2018)1}. TDHF enables us to
study rich and complex physics of low-energy heavy-ion reactions from nucleonic degrees of freedom.

\section{TDHF theory for multinucleon transfer reactions}\label{Sec:MNT}

\subsection{Multinucleon transfer reactions}

\begin{figure*}[t]
\begin{center}
\includegraphics[width=\textwidth]{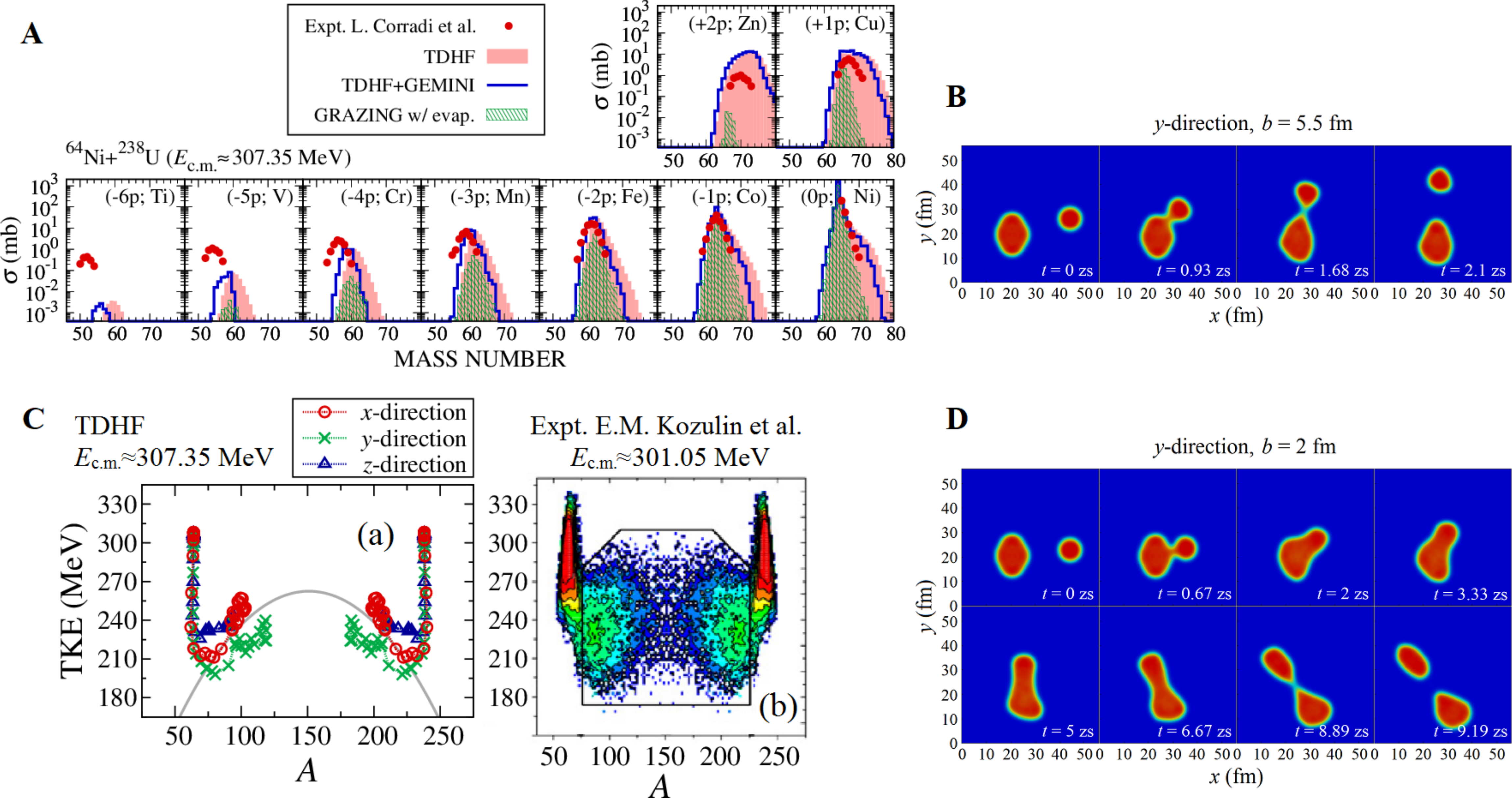}
\end{center}
\caption{
Results of TDHF calculations for the $^{64}$Ni+$^{238}$U reaction at
$E_{\rm c.m.}\approx 307$~MeV are shown as an illustrative example.
\textbf{(A)} Production cross sections for various proton transfer channels
[indicated by ($\pm xp$), where plus (minus) sign corresponds to transfer from
$^{238}$U to $^{64}$Ni (vise versa)] as a function of the mass number of the
reaction products. Red points are the experimental data \cite{Corradi(64Ni+238U)},
red filled histograms are cross sections for primary products by TDHF \cite{KS_KY_Ni-U},
and blue histograms are cross sections for secondary products by TDHF+GEMINI \cite{KS_GEMINI}.
For comparison, cross sections by a semi-classical model, GRAZING \cite{GRAZING-online},
including neutron evaporation effects are also shown by green shaded histograms.
\textbf{(B)} This picture exhibits the time evolution of the density in the reaction plane
in a peripheral collision ($b$\,=\,5.5~\,fm), where MNT processes shown in \textbf{(A)}
take place. Elapsed time in the simulation is indicated in zeptoseconds (1\,zs\,=\,$10^{-21}$\,sec).
\textbf{(C)} Correlation between the total kinetic energy (TKE) and the mass number $A$ of
the outgoing fragments. The left figure (a) shows results of TDHF calculations \cite{KS_KY_Ni-U},
while the right figure (b) shows the experimental data \cite{Kozulin(64Ni+238U)}.
\textbf{(D)} Same as \textbf{(B)}, but for a smaller impact parameter ($b$\,=2\,fm),
where QF processes shown in \textbf{(C)} take place. In the figure, $x$-, $y$-, and
$z$-direction indicate the orientation of deformed $^{238}$U \cite{KS_KY_Ni-U}.
In this way, TDHF can describe both peripheral and damped collisions in a unified way.
Figures were reprinted from Refs.~\cite{KS_KY_Ni-U,KS_GEMINI}.
}
\label{figure}
\end{figure*}

The MNT reaction may be regarded as a non-equilibrium quantum transport process of nucleons during a collision.
With the help of the particle-number projection technique \cite{Projection}, transfer probabilities can be
deduced from a TDHF wave function after collision. In Ref.~\cite{KS_KY_MNT}, a range of reactions at energies
around the Coulomb barrier were studied within TDHF, for which precise experimental data are available
\cite{Corradi(40Ca+124Sn),Corradi(48Ca+124Sn),Corradi(58Ni+208Pb),Szilner(40Ca+208Pb)2}. Those reactions
are characterized by different neutron-to-proton ratio, $N/Z$, and charge product, $Z_{\rm P}Z_{\rm T}$. TDHF
identified two distinct transfer mechanisms: 1) a fast isospin equilibration process and 2) transfer of many nucleons
associated with dynamics of neck formation and its breaking. The neck breaking dynamics emerge at small impact
parameters, especially in reactions with a large $Z_{\rm P}Z_{\rm T}$ ($\gtrsim$\,1600). The latter may
be regarded as a precursor to QF, where the system starts evolving towards the mass equilibrium. Comparisons
with measured cross sections showed that TDHF works fairly well in accuracy comparable to other model predictions.
However, effects of secondary neutron evaporation become substantial, when many protons are transferred.
Combining TDHF with a statistical compound-nucleus deexcitation model, GEMINI++ \cite{GEMINI++}, it became possible to
compute production cross sections after secondary deexcitation processes. With the method, dubbed TDHF+GEMINI,
it was shown that the inclusion of deexcitation effects substantially improves agreement with the experimental data
\cite{KS_GEMINI} [See Fig.~\ref{figure}(A,\,B)]. (For more details, see, e.g., Refs.~\cite{KS_KY_Maruhn,KS_GEMINI}.)

The description is still, however, not perfect, especially for channels where both neutrons and protons are removed.
It would be improved when one includes missing one-body fluctuations into the description (cf. Sec.~\ref{Sec:TDRPA}).
Nevertheless, TDHF may be used, taking advantage of its non-empirical nature, to explore the optimal or novel
reaction mechanism for producing new neutron-rich unstable nuclei, at least qualitatively. Recently, two groups
have also implemented TDHF+GEMINI: production of $N=126$ neutron-rich nuclei was investigated for the
$^{132}$Sn+$^{208}$Pb reaction \cite{Jiang(2018)}; MNT processes in the $^{58}$Ni+$^{124}$Sn reaction
were studied obtaining good agreement with experimental data \cite{Wu(2018)}.

\subsection{Quasifission processes}

To synthesize the heaviest elements and also to explore the predicted island of stability, it is crucial to
establish deep understanding of the main competitive process, the \textit{quasifission}. The QF process
is characterized by a large amount of mass transfer, full energy dissipation, and a long contact time
(typically several to tens of zeptoseconds). The characteristic observables of QF are correlations between
fragment masses with scattering angles, known as mass-angle distribution (MAD), or with the total kinetic
energy (TKE) of outgoing fragments, sometimes called mass-energy distribution (MED) (see, e.g.,
Refs.~\cite{Kozulin(64Ni+238U),Toke(QF1985),Bock(1982),duRietz(2011),duReitz(2013)}). In recent
applications of TDHF, it has been shown that TDHF can quantitatively describe main QF dynamics,
consistent with experimental observations \cite{KS_KY_Ni-U,Wakhle(2014),Umar(2015)1,Hammerton(2015)}
[See Fig.~\ref{figure}(C,\,D)]. An intriguing indication is that in collisions involving an actinide nucleus,
TKE becomes larger (smaller energy dissipation) for tip collisions as compared to side collisions \cite{KS_KY_Ni-U,
Umar(2016)1}, perhaps due to shell effects of $^{208}$Pb \cite{Morjean(2017)}. This may be in contrast
to the naive consideration, where one expects a compact shape at scission for side collisions, which would
result in larger TKE for side collisions.

The revival of interest in the MNT reaction was initiated by the seminal work by Zagrebaev and Greiner
\cite{Zagrebaev(2005),Zagrebaev(2006),Zagrebaev(2007)1,Zagrebaev(2007)2,Zagrebaev(2008)1,
Zagrebaev(2008)2,Zagrebaev(2011),Zagrebaev(2013),Zagrebaev(2014),Zagrebaev(2015)}. In their
work, the importance of shell effects in MNT processes was emphasized. For example, in the $^{238}$U+$^{248}$Cm
reaction, $^{238}$U tends to evolve towards doubly-magic $^{208}$Pb, giving the rest of 30 nucleons
to the heavier partner, resulting in the production of transcurium nuclei as primary reaction products. The
latter process is called \textit{inverse} (anti-symmetrizing) QF (IQF), since the ordinary QF drives the
system towards the mass equilibrium. Clearly, it is essential to correctly include static and dynamic effects
of nuclear shells to describe IQF. The IQF was also observed in TDHF. In Ref.~\cite{Kedziora(2010)},
a typical ``tip-on-side" orientation in collisions of two actinide nuclei, $^{232}$Th+$^{250}$Cf, was
found to result in IQF, where about 15 nucleons in the tip of $^{232}$Th were transferred to the side of
$^{250}$Cf. Another type of IQF was reported in Ref.~\cite{KS_U-Sn}, where complex surface vibration
modes are induced in tip collisions of $^{238}$U+$^{124}$Sn at energies well above the Coulomb barrier,
which resulted in an abrupt development of a neck, forming a smaller subsystem, leading to transfer of around
27 nucleons from $^{124}$Sn to $^{238}$U.
An important message here
is that those IQF processes emerge as a main reaction outcome in TDHF. Since larger effects of fluctuations
and correlations, that give rise to a wider fragment mass distribution, are expected in such damped collisions
\cite{Simenel(review),Ayik(2015)2,Ayik(2017),Ayik(2018)}, novel IQF processes may pave new pathways
to unexplored territories in the nuclear chart far away from the stability.

\section{New techniques and extensions}\label{Sec:extensions}

\subsection{Extraction of macroscopic physical ingredients}

The potential energy surface in a colliding nuclear system is often a key ingredient in phenomenological
approaches. With the constrained Hartree-Fock technique \cite{Ring-Schuck}, one can compute the
adiabatic potential energy in a composite system. However, it differs, in general, from the potential
in the entrance channel of a collision, where incident-energy-dependent dynamic effects come into play.
Such a dynamic potential should be derived from microscopic theories, e.g. TDHF, where dynamic effects
such as shape deformation (or necking), nucleon transfer, and inelastic excitations through the one-body
dissipation mechanism are naturally taken into account. A method named dissipative-dynamics TDHF (DD-TDHF)
\cite{DDTDHF1,DDTDHF2,DDTDHF3} allows us to extract from TDHF not only a dynamic potential, but also
a friction coefficient, based on a mapping of TDHF trajectories onto a set of classical equations of motion.
Density-constrained TDHF (DC-TDHF) \cite{Umar(2006)} is another way to extract a potential. In the latter
approach, the density distribution obtained from TDHF is used as a constraint for Hartree-Fock calculations,
leading to an adiabatic potential along the TDHF dynamical trajectory (see, e.g., Refs.~\cite{Umar(2010),
Umar(2009),Umar(2010)2,Oberacker(2010),Umar(2012)1,Oberacker(2012),Keser(2012),Oberacker(2013),
Simenel(2013)1,Umar(2014),Steinbach(2014),Reinhard(2016),Godbey(2017),Guo(2018)3}, for various applications).
Energy dependence as well as transition from dynamic to adiabatic potentials may be important as well. Later,
the idea of the density constraint was also applied to extract a bare (without density distortion) potential taking
full account of the Pauli exclusion principle, which is named density-constrained frozen Hartree-Fock (DC-FHF)
\cite{DCFHF(2017)}. Such a bare potential can be applied to, e.g., coupled-channels calculations for fusion
reactions \cite{CCFULL,Hagino(2012)}. Indeed, such a connection was established, where excitation energies
of low-lying collective excitations were also determined by TDHF \cite{Simenel(2013)}. This is a remarkable
example of building a bridge between elaborated macroscopic or phenomenological frameworks and those
rooted with microscopic degrees of freedom, that may enhance predictability of the former.

\subsection{Balian-V\'en\'eroni variational principle and TDRPA}\label{Sec:TDRPA}

While there are, of course, limitations inherent in TDHF (e.g., the absence of many-body correlations,
deterministic nature as well as spurious cross-channel couplings due to the single mean-field description
\cite{Ring-Schuck}), some of them can nowadays be overcome with extended approaches. One such
approach can be obtained from the variational principle of Balian and V\'en\'eroni \cite{BV(1981)}.
The Balian-V\'en\'eroni variational principle enables us to control the variational space according to not
only the state in the Schr\"odinger picture, but also to the observable in the Heisenberg picture. For instance,
a variation optimized for Slater determinant and one-body observable derives the TDHF equations \cite{Simenel(review)}.
By extending the variational space for fluctuations of a one-body observable, one obtains a formula that
includes effects of one-body fluctuations on top of the TDHF mean-field trajectory, which may be regarded
as time-dependent RPA (TDRPA). It has recently been shown that the description of the width of
fragment mass distributions in deep-inelastic collisions, which is severely underestimated in TDHF
\cite{Koonin(1977),Davies(1978),Dasso(1979)}, is substantially improved \cite{Broomfield(2009),
Simenel(2011)}, or even quantitatively agrees with experimental data \cite{Williams(2018)}, in TDRPA.
It implies that one-body dissipation and fluctuations described by TDHF and TDRPA, respectively, are the
predominant mechanisms in deep-inelastic collisions. Here, a caution is required: the TDRPA formula in the
present form can only be applied to symmetric reactions \cite{Williams(2018)}. To overcome this difficulty,
one may extend the derivation based on the Balian-V\'en\'eroni variational principle to include higher-order
corrections, or derive a formula generalized for asymmetric reactions from, e.g., the stochastic mean-field
(SMF) theory \cite{Ayik(2008),Lacroix(2014)}, from which one can derive the TDRPA formula in the small
fluctuation limit \cite{Ayik(2008)}. Note that effects of two-body correlations were also indicated by the
time-dependent density-matrix (TDDM) approach \cite{Shun-jin(1985),Gong(1990),Tohyama(2002),
Tohyama(2016),Wen(2018)}.

\subsection{Stochastic extensions}

Another limitation of TDHF lies in its deterministic nature. To correctly describe abundant reaction
outcomes as observed experimentally, one may need to introduce quantal, collisional or thermal
fluctuations that induce bifurcations of dynamics, leading to, in the end, minor processes apart from
the TDHF mean-field trajectory. In the SMF approach, extensively developed by Ayik and his coworkers
\cite{Ayik(2015)2,Ayik(2017),Ayik(2018),Ayik(2008),Lacroix(2014),Ayik(2009),Washiyama(2009)1,
Yilmaz(2011),Yilmaz(2014),Ayik(2015)1,Ayik(2016),Tanimura(2018),Yilmaz(2018)}, initial fluctuations
are introduced in a stochastic manner, which gives rise to an ensemble of final states after dynamic evolutions.
For the MNT study, the SMF description can be cast into the Fokker-Planck equations, where transport (drift
and diffusion) coefficients are determined by the single-particle orbitals in TDHF. In Ref.~\cite{Yilmaz(2018)},
the SMF approach was also applied to the Ni+Ni reaction, showing quantitative agreement with experimental data
\cite{Williams(2018)}, comparable to TDRPA. Recently, Bulgac et al. \cite{Bulgac(2018)} proposed another
way to introduce stochasticity by means of external potentials which are random in both space and time. The
method was applied to describe induced fission processes, with a simplified orbital-free approach, getting a
sufficient width of the fission fragment mass distribution \cite{Bulgac(2018)}. Further applications of those
approaches may be promising for exploring optimal reactions, especially for producing the most exotic isotopes,
where correct description of rare processes may be essential.

\subsection{Inclusion of pairing}

In spite of the known great importance of pairing in nuclear structure studies, the effect of pairing on nuclear
reaction dynamics has rarely been investigated so far. While the BCS approximation has been employed to date
\cite{Scamps(2012),Scamps(2013),Ebata(2014)2,Ebata(2015)}, inclusion of dynamic effects of pairing became
possible only quite recently with top-tier supercomputers. One may naively expect that pairing in the nucleus
is so fragile and it would only affect tunneling phenomena below the Coulomb barrier, where clear effects of
nucleon-nucleon correlations were indeed observed experimentally \cite{Evers(2011),Corradi(2011),Montanari(2014),
Montanari(2016)}. To the contrary, Magierski et al. \cite{MSW(2017),SMW(2017),SWM(2017)} found unexpectedly
large effects associated with the relative phase (or gauge angles) of colliding superfluid nuclei on the reaction
outcomes, such as TKE and fusion cross section. In this context, experimental fusion cross sections were analyzed,
which indicates fusion barrier width may be increased by the superfluid effect \cite{Scamps(2018)1}. In the work
by Magierski et al., a local formulation of superfluid TDDFT, known as time-dependent superfluid local density
approximation (TDSLDA) (see, e.g., Refs.~\cite{Bulgac(2012),Bulgac(2013),Magierski(2016),Magierski(2018)}),
was employed, which is formally similar to the time-dependent Hartree-Fock-Bogoliubov (TDHFB) approach \cite{Ring-Schuck}.
The TDSLDA has been extensively developed by Bulgac and his coworkers, which was successfully applied to the unitary
Fermi gas \cite{Bulgac(2009),Bulgac(2011),Bulgac(2012)2,Bulgac(2014),Wlazlowski(2015),Wlazlowski(2018)},
finite nuclei \cite{Stetcu(2011),Stetcu(2015),Bulgac(2016),Magierski(2017),Bulgac(2018)2,Stetcu(2018)},
as well as interiors of neutron stars \cite{Wlazlowski(2016)}. Hashimoto has developed a Gogny TDHFB code
that works with 2D-harmonic-oscillator plus 1D-Lagrange-mesh hybrid basis \cite{Hashimoto(2013)}, which
could be applied for head-on collisions \cite{Hashimoto(2016),Scamps(2017)}. Qualitatively the same effects
of the relative phase as reported in Ref.~\cite{MSW(2017)} were observed with their code, although smaller
in magnitude for heavier systems \cite{Hashimoto(2018)}---further investigations are required. We have just
entered a new era in which dynamic effects of pairing can be fully incorporated with microscopic nuclear
dynamics studies. With the aforementioned sophisticated approaches with the usage of top-tier supercomputers,
the dynamic effects of pairing in MNT and QF processes will be unveiled in the near future.

\subsection{Remarks on other theoretical approaches}

There are a bunch of models developed so far for the MNT study, such as the semi-classical model, like GRAZING
\cite{GRAZING-online,GRAZING1,GRAZING2,Dasso(1994),Dasso(1995),Yanez(2015),GRAZING-code}
or CWKB \cite{Corradi(48Ca+124Sn),Corradi(58Ni+208Pb),Szilner(40Ca+208Pb)2,CWKB}, the dinuclear system
(DNS) model \cite{Adamian(2005),Penionzhkevich(2005),Penionzhkevich(2006),Feng(2009),Adamian(2010)1,
Adamian(2010)2,Adamian(2010)3,Mun(2014),Mun(2015),Zhu(2015),Zhu(2016),Feng(2017),Zhu(2017)1,Li(2017),
Zhu(2017)2}, the improved quantum molecular dynamics (ImQMD) model \cite{Zhao(2015),Li(2016),Wang(2016),Zhao(2016)},
and the Langevin model \cite{Zagrebaev(2005),Zagrebaev(2006),Zagrebaev(2007)1,Zagrebaev(2007)2,Zagrebaev(2008)1,
Zagrebaev(2008)2,Zagrebaev(2011),Zagrebaev(2013),Zagrebaev(2014),Zagrebaev(2015),Karpov(2017),Saiko(2018)}.
There are pros and cons in those approaches. For instance, on the one hand the semi-classical model can describe
successive transfer processes at peripheral collisions, on the other hand it misses deep-inelastic components at
small impact parameters. In contrast, the DNS model is capable of describing the dynamic evolution of a composite
system, according to a master equation with a potential energy (including shell effects) for mass asymmetry and deformation
of the subsystem, that gives rise to a probability distribution for massive nucleon transfer as well as fusion. By construction,
however, the latter model assumes a formation of a `di-nucleus' in a potential pocket, and thus it misses (quasi)elastic
components at large impact parameters. Actually, a simple addition of those model predictions, DNS+GRAZING,
was considered in the literature \cite{Welsh(2017),Wen(2017)}. The ImQMD model is applicable for both peripheral
and damped collisions. By taking into account stochastic nucleon-nucleon collisions, it provides distributions of
observables accumulated in a number of simulations. To the author's knowledge, however, the spin-orbit interaction
has been neglected in ImQMD, that prevents a proper description of shell effects in MNT and QF processes. The
elaborated Langevin model \cite{Karpov(2017)} may be promising in describing complex processes of MNT, QF
and fusion, in a unified way. The model contains, however, various parameters that have to be tuned carefully
to reproduce available experimental data \cite{Karpov(2017)}.

A possible future task is to non-empirically determine model ingredients based on microscopic theories:
potential energy surface with respect to an arbitrary set of variables, drift, diffusion and friction coefficients, etc.,
can be derived from the microscopic approaches, that may lead to, e.g., kind of TDHF+Langevin approach.

\section{Summary and prospect}\label{Sec:summary}

Production of neutron-rich heavy nuclei is listed as one of the high-priority important subjects of nuclear
science today. In this contribution, recent advances in the MNT study, especially those based on microscopic
dynamic theories, have been briefly reviewed. The traditional TDHF approach has shown remarkable successes in
describing a variety of phenomena in nuclear systems. It has been shown that TDHF is capable of describing the main
reaction channels, not only MNT processes in peripheral collisions, but also deep-inelastic and QF processes in damped
collisions, placing it as a good starting point for building a fully microscopic theory for low-energy heavy-ion reactions.
Moreover, the inclusion of one-body fluctuations significantly improves quantitative agreements of the width of fragment
mass distributions with experimental data. With the use of top-tier supercomputers, studies of dynamic effects of
pairing have just been started. One of the great advantages of microscopic approaches is its non-empirical
nature---one may explore novel pathways towards the dreamed of production of yet-unknown neutron-rich nuclei.
Another possible future direction is to use microscopic approaches to determine phenomenological parameters in
other theoretical models. It may be the time to combine knowledge obtained with different approaches to construct
the most reliable framework to lead future experiments to the successful production of extremely neutron-rich nuclei,
the farthest from the continent of stability.

\begin{acknowledgements}
The author acknowledges P.D.~Stevenson, J.A.~Maruhn, C.~Simenel, Lu Guo, and A.S.~Umar
for careful reading of the manuscript and for providing valuable comments on this article.
\end{acknowledgements}

\end{document}